# A Graph Theoretic Approach for Optimizing Key Pre-distribution in Wireless SensorNetworks[*]

Aldar C-F. Chan
School of Computing
National University of Singapore
email: aldar@comp.nus.edu.sg

July 1, 2018

**Abstract**

Finding an optimal key assignment (subject to given constraints) for a key predistribution scheme in wireless sensor networks is a difficult task. Hence, most of the practical schemes are based on probabilistic key assignment, which leads to sub-optimal schemes requiring key storage linear in the total number of nodes. A graph theoretic framework is introduced to study the fundamental tradeoffs between key storage, average key path length (directly related to the battery consumption) and resilience (to compromised nodes) of key predistribution schemes for wireless sensor networks. Based on the proposed framework, a lower bound on key storage is derived for a given average key path length. An upper bound on the compromising probability is also given. This framework also leads to the design of key assignment schemes with a storage complexity of the same order as the lower bound.

## 1 Introduction

A sensor network is made up of low-cost nodes operating on battery power. Hence, computation and energy consumption are major constraints for the design of sensor networks [3]. Similar to key distribution schemes with an online trusted third party server [23, 24], the main design goal of key management in a sensor network is for entity authentication or initial trust establishment, that is, to allow the sensor nodes to identify each other or to distinguish between insiders and outsiders (adversarial nodes) during deployment.

Since the nodes in a sensor network usually belong to the same owner, it is thus fairly reasonable to assume all the nodes have prior contact with a common entity or server before deployment; nevertheless, using an online server as in [23, 24] is considered as impractical for sensor networks. As a result, to minimize the consumption of resources like computation and energy in order to fulfill the low cost constraint, key predistribution schemes (introduced in [19, 22]) have emerged as the de facto method for achieving initial trust in a sensor network despite that a considerable among of key storage is usually needed [1, 2, 4, 10–13, 16–18]. Note that the family of probabilistic key predistribution schemes [4, 10, 11, 13, 17, 18] requires key storage which is a linear function of the number of nodes in the network. In other words, storage is traded off for reduced computation and longer battery lifetime. Although this initial trust also provides private channels among sensor nodes to establish session keys for secret and authenticated communication, entity authentication is still the main goal of key predistribution schemes (KPS) for any sensor network.

---

[*]An earlier version of this paper appeared in the proceeding of ICST WiOpt'09.

In KPS, each node is preloaded with some set of secret keys into its key ring before deployment and it is assumed that only the owner or nodes belonging to the same owner would know any of these keys. Of course, a compromised node would also know part of these keys; so it is one of the design criteria of KPS to minimize the impact of a compromised key ring. When two nodes find they share one or more common keys, they will recognize each other as belonging to the same owner and thus trust each other. Note that this trust is mutual in the sense that when node A trusts node B, node B should also trust node A. This trust could also be called a keying relationship between A and B as in [21]. When two nodes, A and B, share no common key, they cannot identify each other immediately, but if both of them possess established trust with a third node C, of course through shared keys again, they can establish trust through C and set up a new key between them to be used as a sign of trust afterwards. We call the path A-C-B an authentication or acquaintance chain between A and B. Although misleading, this trust establishment process is commonly known as path key establishment in the context of sensor network key management since [13]. It is straightforward to see how this trust establishment can be extended to a longer acquaintance chain, for instance, A-C-D-E-...- B. This is analogous to the authentication chain [6, 15] in Public Key Infrastructure (PKI) [8], but they differ in that the trust of the certificate chain in PKI is unidirectional while the trust relationship between two nodes connected by an acquaintance chain in KPS is mutual; in PKI, when A trusts that a public key belongs to B through some authentication path in PKI, it is not necessary that B would trust the authenticity of A's public key.

With this background, we could consider the design objective of any KPS for a sensor network as finding a key assignment method allowing any two sensor nodes to authenticate each other through a reasonable-length authentication chain with reasonable resource consumption, key storage requirement and security, dependant on the application scenarios. Various design variations could be came up with by varying the tradeoff between any pair of the desired properties or parameters of a KPS, including key storage requirement, network connectivity, energy consumption, resilience to dead nodes and security against compromised nodes.

When discussing entity authentication, it is natural to represent the keying or trust relationships between nodes by a graph, just like the certificate graph in PKI [6, 21]. We call this type of graph a trust graph (t-graph). In a t-graph $G$, the sensor nodes are represented as vertices in $G$ and an edge exists between two vertices only if the corresponding sensor nodes have direct established trust (i.e. they share some common keys). The t-graph is indeed used in many existing schemes on sensor network key management [4, 13] to represent the keying relationships between nodes. However, the existing work merely uses a t-graph for representation purposes to calculate the probability of connectivity of the actual network. We especially study the tradeoff between storage and energy consumption for full connectivity. They have overlooked the power of a t-graph as a bridge between the design and analysis of sensor network key management and the result-rich graph theory. In this paper, we show how a t-graph can be leveraged to demonstrate the tradeoff between the properties of KPS for sensor networks, unify the existing schemes under this framework, and give design insight based on graph theoretic results, in particular those about graph diameters and intersection graphs. We also give a lower bound on key storage based on the proposed framework. This lower bound is generic in the sense that it applies to all key predistribution schemes and the graph model in this paper is merely a means to derive this lower bound but poses no restriction on the key assignment in the actual KPS.

The main contribution of this paper is the graph theoretical framework proposed to study key assignment methods for key predistribution in wireless sensor networks. Such a realization provides a unifying view for predistribution schemes and allows the derivation of key storage lower bound and compromising probability upper bound for a given constraint on the average key path length (which is directly related to energy consumption for bootstrapping a sensor network). The proposed framework also links results in graph theory to the design of key assignment schemes. Through such link, we provide a number of near optimal key assignment methods.

The organization of this paper is as follows. The definition of a t-graph is given in the next section together with a discussion of the physical connectivity graph of a sensor network. Then, it will be shown how the properties or parameter of a KPS is related to the parameters of a t-graph in Section 3. In Section 4, the tradeoff between these parameters is discussed. After that, a number of near-optimal key assignment methods are given in Section 5.

## 2 Trust Graph

In a t-graph or the associated key graph (which can be used to represent any KPS), each node in a sensor network is represented as a vertex. In rigorous definitions, there are two main differences between a t-graph and its key graph. First, each edge in a t-graph represents established trust between the two end vertices while each edge in a key graph represents a key shared by the end vertices. Hence, multiple edges between two adjacent vertices are possible in a key graph but not in a t-graph since two sensor nodes may share more than one key but the trust relationship is binary (that is, two sensor nodes could mutually trust each other or not). Second, a t-graph is an unlabeled graph while the edges of a key graph are labeled by keys or key indices. The definitions of a t-graph and its key graph are given below.

**Definition 1 [t-Graph]** *Given a set of sensor nodes $\mathcal{S} = \{s_1, s_2, \ldots, s_n\}$, a t-graph $G$ is an unlabeled, non-weighted graph with a vertex set $V(G) = \mathcal{S}$ such that $(s_i, s_j) \in E(G)$ (the edge set of $G$) where $i \neq j$ if and only if $s_i$ and $s_j$ have direct established trust, that is, share some common keys.*

**Definition 2 [Key Graph]** *Given a set of sensor nodes $\mathcal{S} = \{s_1, s_2, \ldots, s_n\}$ and a key set $\mathcal{K} = \{k_1, k_2, \ldots, k_t\}$, a key graph $G$ is an edge-labeled or edge-colored, non-weighted graph with a vertex set $V(G) = \mathcal{S}$, a label set equal to $\mathcal{K}$ (or $[1,t]$) and an edge set $E(G) \subseteq \mathcal{S} \times \mathcal{S} \times \mathcal{K}$ such that $(s_i, s_j, k_r) \in E(G)$ where $i \neq j$ if and only if $s_i$ and $s_j$ share a common key $k_r$.*

It should be noted that $s_i$ is just some kind of label for a sensor node and it could be of any format, not merely the identity of a sensor node. If we label a sensor node by the set of keys it holds, the resulting t-graph is an intersection graph [20].[1]

For the sake of clarity, in this paper, we consider a simple key assignment wherein each edge in a t-graph is associated with a key. That is, the key graph can simply be obtained from a t-graph by assigning key labels to it. Consequently, the problem of designing optimal key assignments reduces to finding an optimal t-graph (with desired tradeoffs) for a given application scenario, and then assigning keys to links in the t-graph.

The degree of a node in a t-graph, more precisely, the total number of key labels of all the edges ending at a node in a key graph, tells the required key storage of that node. Nevertheless, in most cases, the degree of a node in the t-graph gives a good indication of the key storage requirement. In any proper KPS, both the t-graph and key graph should be connected, that is, any vertex in the graph should be reachable by another. As usual, the distance between any two vertices, denoted by $d(s_i, s_j)$, is the number of edges/hops traversed by the shortest path between $s_i$ and $s_j$. If two nodes are not connected, the corresponding distance is $\infty$, which normally should not happen in a proper KPS. Suppose a t-graph $G$ has $n$ vertices, the mean distance $\overline{d}$ and the diameter $D$ of $G$ are defined as follows:

$$\overline{d} = \tfrac{1}{\binom{n}{2}} \sum_{s_i, s_j \in V(G)} d(s_i, s_j),$$
$$D = \max_{s_i, s_j \in V(G)} d(s_i, s_j).$$

The mean distance and diameter are explicitly discussed here because they are related to the length of the acquaintance chains in a particular t-graph, which are in turn related to the energy consumption of establishing trust indirectly between two nodes through an acquaintance chain. The details will be discussed in Section 3.

## 2.1 Physical Connectivity and Deployed t-Graph

Recall that KPS aims at allowing physical neighboring nodes (within the radio range of each other) to identify each other and establish trusted links. If the actual deployment topology is known, we just need to make a t-graph the same as the physical connection topology of the sensor network. The physical connection could be best represented by a physical connectivity graph (or physical graph for short) in which sensors are represented by vertices and an edge exists between two vertices if and only if they are within the radio range of each other. A physical graph is the actual physical topology of a sensor network in deployment. In other words, an edge in the physical graph represents that the two end nodes of the edge are within reach of each other's radio range and communicate directly without relying on others to relay messages.

---

[1] It has been shown that each graph is an intersection graph [20]. An explicitly vertex-labeled intersection graph is meant here.

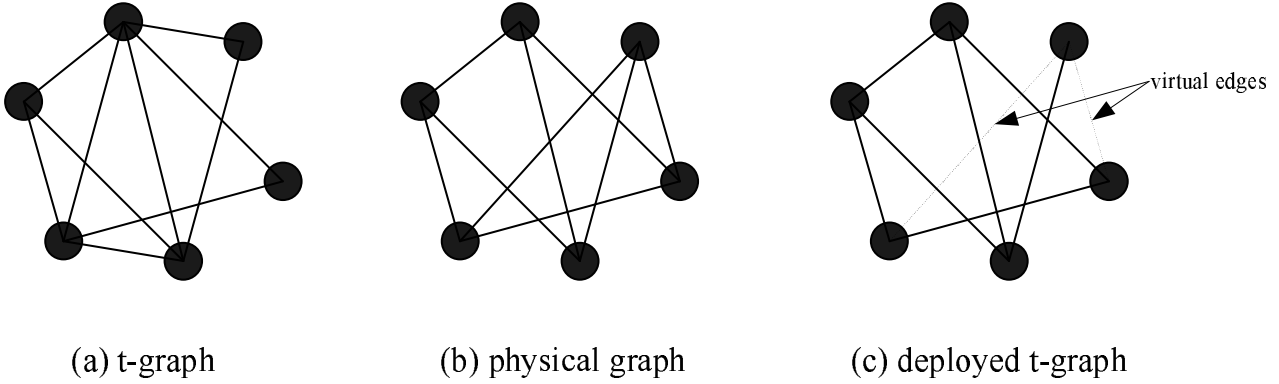

(a) t-graph  (b) physical graph  (c) deployed t-graph

Figure 1: Obtaining a deployed t-graph (representing the trusted links that can be established after bootstrap) from the original t-graph (representing the established trust relationships among sensor nodes) and a physical graph (representing the physical link connectivity in the actual deployed network).

We should distinguish a t-graph or key graph from a physical graph of a sensor network; while an edge in the latter represents a real physical link between two neighboring sensor nodes, an edge in a t-graph (key graph) is just a logical link. In order to set up or boot-strap a secure network, only physically neighboring nodes (connected by an edge in the physical graph) having mutual trust (represented by an edge in the t-graph) can establish a trusted link. That is, even though two nodes are neighbors in the t-graph, the bootstrapped initial trust between them would not be useful if they are not neighbors in the physical graph. On the hand, two neighboring nodes within the radio range of each other still cannot authenticate or verify the identity of the neighbor if they share no initial secret. If we want to see how many trusted links are established in the actual network or equivalently how many edges in the t-graph are effective for setting up trusted links in the actual network, we could obtain a deployed t-graph by computing the graph-intersection of the t-graph and the physical graph. In a deployed t-graph, only edges actually used to set up trusted links in the network are shown. However, an edge in the physical graph not overlapped with an edge in the t-graph is shown as a dashed line, called a virtual edge. A virtual edge corresponds to a physical link between two sensor nodes in physical proximity having no mutual trust. Depicted in Figure 1 shows how a deployed t-graph can be obtained by overlapping the original t-graph with a physical graph.

Additional trusted links can be established from these virtual edges as the nodes in question could authenticate each other through an acquaintance chain, thus adding edges to the deployed t-graph. The procedure is as follows: Suppose $s_i$ and $s_j$ (in physical proximity) wish to authenticate each other through an acquaintance chain $s_i - s_a - s_b - s_j$. Note that each neighboring pairs in this chain need to be physical neighbors with established trust. $s_i$ randomly picks a key $k$, encrypts $k$ using the key it shares with $s_a$, and passes it to $s_a$. Then $s_a$ retrieves $k$ and encrypts it using the key it shares with $s_b$ and passes it to $s_b$. This process continues until $s_j$ is reached. Finally, $s_j$ picks another key $k'$ and encrypts it using $k$ and sends it back to $s_i$. The key $k'$ or a combination of $k$ and $k'$ would be used as a sign of trust established indirectly between $s_i$ and $s_j$. The last step is necessary for key confirmation. This process in essence converts virtual edges into trusted link in the deployed t-graph. How many successful conversion can be made depends on the acceptable maximum length of the acquaintance chain. If there is no restriction on this length, for a network with proper key assignment, all virtual edges or physical links can be turned into trusted links.

The goal of any KPS for a sensor network is to convert as many virtual edges as possible into real edges in the deployed t-graph subject to resource constraints like energy consumption (which poses constraints on the maximum length of the acquaintance chain). The energy consumption cost of adding one such edge would depends on the length of the acquaintance chain which is related to the mean distance and the diameter of the deployed t-graph. Note that for any pair of nodes, the actual acquaintance chain in the deployed network is likely to be different from that in the original t-graph, and only nodes connected by a multi-hop path in the deployed t-graph could identify each other through an acquaintance chain.

It should be noted that a virtual edge is not an authenticated link and one of its ends could possibly be a pre-deployed adversary node. If we insist on using these virtual edges as channels to relay messages for indirect trust establishment, a shorter physical path for the acquaintance chain is possible but the reliability of this physical path is in doubt. Even worse is a long acquaintance chain including just one untrusted edge would render the whole chain insecure, that is, the two end nodes of the chain are unsuccessful in verifying the identity of each other. Hence, this approach is feasible only for scenarios requiring a low level of security on entity authentication.[2]

Since some of the sensor nodes may die before or during deployment, the resulting physical graph and thus the deployed t-graph could have a smaller number of vertices than that in the original t-graph. Similarly, it is impossible to achieve perfect connectivity in the physical graph due to limited radio range of each sensor node; consequently, the number of edges in the immediate deployed t-graph could be significantly less than that in the original t-graph. These two phenomena are equivalent to deleting vertices and edges respectively in the original t-graph to obtain the deployed one, causing disconnection of two originally connected vertices. These events of disconnection lead to an increase in the diameter and mean distance of the deployed t-graph compared to the original t-graph, and if a significant number of them occur, the deployed graph could become disconnected even though the original t-graph is connected. The length of the acquaintance chain between any two non-neighboring vertices increases accordingly. For instance, the two virtual edges shown in Figure. 1 need a 3-hop chain to authenticate each other whereas in the original t-graph a 2-hop chain exists. As a result, node degrees in a t-graph (and key storage requirement as a consequence) needs to be significantly higher than the number of physical neighbors in the actual network in order to tolerate edge and vertex deletion as the physical topology and the resulting

---

[2]In these scenarios, the number of sensor nodes is usually insignificant and key storage is thus not a serious constraint. As a result, this approach may not have any application.

physical graph are unknown prior to deployment in most cases.

## 2.2 Physical Connectivity Model

As can be seen in the last section, any KPS design needs to suit a particular physical connectivity model. If the physical connectivity of the actual deployment is known, the t-graph for optimal key assignment is simply the given physical connectivity graph. This paper considers the random graph model in which each node has on average $b$ neighbors (including both dead and living nodes) and a fraction $p_{die}$ of nodes die before deployment. The resulting physical graph has $n(1 - p_{die})$ vertices on average and the probability that an edge exists between a pair of nodes is given by

$$p_{link} = (1 - p_{die})^2 \frac{b}{n-1}.$$

It can be shown that, in this kind of physical connectivity model, the parameters (like average node degree and average distance) of the resulting deployed trust graph differs from that of the original t-graph by a constant factor where the constant is function of $p_{link}$ only. As a result, for the sake of clarity, our discussions in this paper will be mainly based on the t-graph rather than the actual deployed t-graph.

## 3 Relations between KPS Parameters and the Properties of t-graph

This section shows how the KPS design parameters are related to the properties and parameters of the t-graph. In the following discussion, a t-graph is denoted by $G_T$, its key graph by $G_K$, the deployed t-graph by $G_{DT}$, and a physical graph by $G_P$. To distinguish the parameters of a t-graph from that in a key graph or a deployed t-graph/key graph, the parameters are attached with a subscript of $T$, $K$ or $DT/DK$ respectively. For example, the diameters of a t-graph and a deployed t-graph are denoted by $D_T$ and $D_{DT}$ respectively in the following discussion.

### 3.1 Storage

The key storage requirement $l(s_i)$ of a node $s_i$ is equal to the total number of different key labels on all the edges (or equivalently the total number of edges) ending at that node in the key graph $G_K$. Similarly, the total number of keys used by the network $L$ is equal to the total number of different key labels on all the edges of the key graph.

$$l(s_i) = |\bigcup_{s_j, \exists (s_i, s_j, k_r) \in G_K} \{k_r\}|,$$
$$L = |\bigcup_{s_i, s_j, \exists (s_i, s_j, k_r) \in G_K} \{k_r\}|.$$

If we consider a simple key assignment method in which each edge in the t-graph is assigned a single distinct key, the storage at a node $s_i$ is simply $l(s_i) = deg(s_i)$ (node degree of $s_i$) and the total number of keys is $L = E(G_T)$.

## 3.2 Connectivity

We consider the connectivity of the deployed t-graph $G_{DT}$ obtained by overlapping a t-graph $G_T$ with a physical graph $G_P$. The probability $p_c$ that any two nodes, say, $s_i$ and $s_j$, are connected by an edge in $G_{DT}$ is given by:

$$\begin{aligned}
p_c &= p(s_i, s_j) \\
&= Pr[\{(s_i, s_j) \in E(G_P)\} \wedge \{(s_i, s_j) \in E(G_T)\}] \\
&= p_{link} \frac{|E(G_T)|}{\binom{n}{2}} \\
&= (1 - p_{die})^2 \frac{b}{n-1} \frac{\frac{1}{2}\sum_{s_i \in V(G_T)} deg(s_i)}{\frac{1}{2}n(n-1)} \\
&= (1 - p_{die})^2 \frac{b}{n-1} \frac{\sum_{s_i \in V(G_T)} deg(s_i)}{n(n-1)}
\end{aligned}$$

Note that $p_c$ in the deployed trust graph and its counterpart in the original t-graph differ by a factor of $p_{link}$ as mentioned in the last section. Suppose the minimum and maximum degrees of $G_T$ are denoted by $\theta_{min}$ and $\theta_{max}$ respectively. Since $n\theta_{min} \leq \sum_{s_i \in V(G_T)} deg(s_i) \leq n\theta_{max}$, we have the edge connectivity probability $p_c$ given by

$$(1 - p_{die})^2 \frac{b\theta_{min}}{(n-1)^2} \leq p_c \leq (1 - p_{die})^2 \frac{b\theta_{max}}{(n-1)^2}.$$

This edge connectivity probability determines whether the deployed t-graph remains connected. As a rough estimate, the required minimum degree of the original t-graph could be determined using $p_{link}$, which is summarized as below.

**Theorem 1** *To ensure that the deployed t-graph is connected (with reasonably high probability) for a given $p_{link}$, the minimum degree of the original t-graph $G_T$ is $(1 - p_{link})|E(G_T)| + 1$.*

**Proof** The probability that an edge in $G_T$ is deleted is given by $(1 - p_{link})$, then the expected number of edges deleted from $G_T$ is given by $a = (1 - p_{link})|E(G_T)|$. To ensure that the deployed graph is connected after deleting $a$ edges, $G_T$ must be $a + 1$ edge-connected, thus implying a minimum degree of $a + 1$. ∎

A more accurate estimate for the required degree could be determined using the Erdös and Rényi random graph [25], but such an extension will not be discussed in this paper.

## 3.3 Energy Consumption

Suppose two nodes $s_i, s_j$ (connected by a virtual edge in the deployed t-graph) establish a new edge through an acquaintance chain, the energy consumption is given by

$$W(s_i, s_j) = d_{DT}(s_i, s_j) + 1$$

where $d_{DT}(s_i, s_j)$ is the distance between $s_i$ and $s_j$ in the deployed t-graph. Note that this equation also applies to neighboring nodes with $d_{DT}(s_i, s_j) = 1$. Overall, the maximum energy consumption $W_{max}$ over all node pairs is equal to the diameter of the deployed t-graph, that is,

$$W_{max} = D_{DT} + 1$$

Recall that $G_P$ is the physical graph and we wish to establish trusted links on all edges of $G_P$. The average energy consumption per node pair $\overline{W}$ is given by

$$\overline{W} = 1 + \frac{1}{|E(G_P)|} \sum_{(s_i,s_j) \in E(G_P)} d_{DT}(s_i, s_j).$$

The above energy consumption indices have a unit of average energy per 1-hop transmission and are expressed in terms of the diameter and distances in the deployed t-graph after vertices and edges are deleted from the original t-graph; they do not help in the design of KPS for a sensor network in which we can only control the diameter and distances in the original t-graph. In fact, without the knowledge of the connectivity of the original t-graph, there is not much we can tell about the diameter or distance increase in the deployed t-graph. For general t-graphs, the following theorems by Chung et. al. [5] could be helpful.[3] Nevertheless, without knowledge of the original t-graph, the bound about deletion of vertices does not do any better than the loose bound $D_{DT} \leq (n(1 - p_{die}) - 1)$. However, if we assume the fraction of dead nodes is negligible, the bound about edge deletion could be useful.

**Theorem 2** Diameter Increase After Edge Deletion *[5]. For a $t + 1$ edge-connected graph $G$ with diameter $D(G)$ where $t \geq 1$, if $t$ edges are deleted from $G$, the resulting graph $G'$ has the following diameter upper bound:*

$$D(G') \leq (t+1)D(G) + t$$

**Theorem 3** Diameter Increase After Vertex Deletion *[5]. For a $\lambda$ vertex-connected graph $G$ having $n$ vertices with diameter $D(G)$, for $t < \lambda$, if $t$ vertices are deleted from $G$, the resulting graph $G'$ has the following diameter upper bound:*

$$D(G') \leq \lfloor \frac{n - t - 2}{\lambda - t} \rfloor + 1$$

Although it is almost impossible to find a nice upper bound on the diameter $D_{DT}$ and average distance $\overline{d}_{DT}$ of a deployed t-graph $G_{DT}$ obtained from any general t-graph, in the following, we will characterize a t-graph with its node degrees, diameter and mean distance, and the minimum number of disjoint paths between any pair of non-adjacent vertices with a view to obtaining some bounds for the energy consumption indices. The result is given by the following claim:

**Theorem 4** *Suppose a t-graph $G_T$ with $n$ vertices, diameter $D_T$, average distance $\overline{d}_T$ and minimum degree $\theta_{min}$, and each pair of vertices in $G_T$ have at least $f$ disjoint shortest-path acquaintance chains. Given the diameter $D_{DT}$ of the resulting deployed t-graph $G_{DT}$ (obtained by overlapping with a physical graph), the average distance $\overline{d}_{DT}$ of $G_{DT}$ taken over all physical neighbors is bounded above by the following:*

$$\overline{d}_{DT} \leq \overline{d}_T + (D_{DT} - 2)\left(1 - \frac{\theta_{min}}{n-1}\right) \cdot P$$

---
[3]A $t$ edge-connected graph is a connected graph such that if less than $t$ edges are removed, the graph remains connected. Similarly, a $t$ vertex-connected graph is a connected graph such that is less than $t$ vertices are removed, the graph remains connected.

where $P = \left[1 - (1-p_{die})^{D_T-1} \left(\frac{b}{n-1}\right)^{D_T}\right]^f$.

**Proof** Suppose we pick two node $s_i$ and $s_j$ which are physical neighbors. Let their distance in $G_T$ be $d' = d(s_i, s_j) \leq D_T$. Assume there are $f' \geq f$ disjoint path of this length. There are $d'$ edges and $(d'-1)$ vertices between $s_i$ and $s_j$ in $G_T$.

$$Pr[\text{1 path remains connected in } G_{DT}] = (1-p_{die})^{d'-1} \left(\frac{b}{n-1}\right)^{d'}$$

$$Pr[\text{All } f' \text{ paths are disconnected in } G_{DT}] = \left[1 - (1-p_{die})^{d'-1} \left(\frac{b}{n-1}\right)^{d'}\right]^{f'}$$

$$\leq \left[1 - (1-p_{die})^{D_T-1} \left(\frac{b}{n-1}\right)^{D_T}\right]^{f}$$

The probability of disconnection between $s_i$ and $s_j$ is:

$$p_{dc} = Pr[s_i, s_j \text{ are not neighbors in } G_T \text{ and all } f' \text{ paths are disconnected}]$$

$$\leq \left(1 - \frac{\theta_{min}}{n-1}\right) \left[1 - (1-p_{die})^{D_T-1} \left(\frac{b}{n-1}\right)^{D_T}\right]^{f}.$$

Note that $d' \geq 2$ and the new distance between $s_i, s_j$ should be at most $D_{DT}$ (diameter). Therefore the distance increase is given by $\Delta d(s_i, s_j) \leq D_{DT} - 2$.

Since whether two nodes are physical neighbors are randomly picked, we could take the mean distance $\overline{d}_T$ in $G_T$ as the mean distance of the subset of pairs connected by physical links. As a result, the new mean distance $\overline{d}_{DT}$ in $G_{DT}$ is given by:

$$\overline{d}_{DT} = \overline{d}_T + \frac{1}{|E(G_P)|} \sum_{s_i, s_j, \exists (s_i, s_j) \in E(G_P)} p_{dc} \Delta d(s_i, s_j).$$

This concludes the proof. ∎

Depending on the scenario in question, we could use the bound in Chung's theorem (Theorem 2) or the very loose bound $n(1-p_{die}) - 1$ to substitute $D_{DT}$ in this equation. Note also that the average energy consumption $\overline{W}$ is given by

$$\overline{W} \leq 1 + \overline{d}_T + (D_{DT} - 2)\left(1 - \frac{\theta_{min}}{n-1}\right) \times \left[1 - (1-p_{die})^{D_T-1} \left(\frac{b}{n-1}\right)^{D_T}\right]^{f}.$$

From this equation, we can easily see that as the mean distance $\overline{d}_T$ or diameter $D_T$ of the original PR graph drops, the average energy consumption $\overline{W}$ drops, and as the minimum number $f$ of disjoint paths or the minimum degree $\theta_{min}$ of the original graph increases, the average energy consumption decreases (which is expected). If Chung's theorem is used, for $D_T = 2$, there is almost a linear relationship between $\overline{W}$ and vertex degree $\theta$ of a regular graph in which $\theta_{max} = \theta_{min} = \theta$.

Although the connectivity discussed is Section 3.2 can be considered as a special instance of neighborhood connectivity with $h = 1$, we should distinguish between the two as the former refers to the whether a t-graph remains connected after edges and vertices are deleted from it due to limited radio range and dead nodes in the actual deployment, and without the energy consumption constraint, all edges in a physical graph could be connected.

## 4  Storage Lower Bound

Without loss of generality, we could assume that each edge in a t-graph is assigned a unique, distinct key, and as a result, the degree of a vertex in the t-graph is the key storage requirement at it. The following theorems relate the diameter and mean distance of a t-graph with its node degree. This also states the storage lower bound for any key predistribution scheme for a given key path length that can be tolerated. These lower bounds apply to all key predistribution schemes and demonstrate the trade-off between storage and energy consumption for trust establishments in a sensor network.

The following two theorems give the minimum initial secret key storage needed for bootstrapping a sensor network subject to a certain maximum length restriction on the acquaintance chain.

**Theorem 5** *Given a t-graph $G_T$ having $n$ vertices with diameter $D_T$, for the maximum node degree $\theta_{max} > 2$, the following holds:*

$$\theta_{max} \geq 1 + \sqrt[D_T]{n}$$

**Proof**  Randomly pick a node $s_i$, there are at most $\theta_{max}$ nodes 1-hop away. At 2 hops away, there are at most $\theta_{max}(\theta_{max} - 1)$. At $i$ hops away, there are at most $\theta_{max}(\theta_{max} - 1)^{i-1}$. Hence, the total number of nodes within $D_T$ hops away from $s_i$ is given by:

$$1 + \theta_{max} + \theta_{max}(\theta_{max} - 1) + \ldots + \theta_{max}(\theta_{max} - 1)^{D_T - 1}.$$

Since any two nodes in $G_T$ should be no more that $D_T$ hops apart, the above number should be greater than $n$, that is,

$$\begin{aligned}
n &\leq 1 + \theta_{max} + \theta_{max}(\theta_{max} - 1) + \ldots + \theta_{max}(\theta_{max} - 1)^{D_T - 1} \\
&= 1 + \frac{\theta_{max}[(\theta_{max} - 1)^{D_T} - 1]}{(\theta_{max} - 1) - 1} \\
&= \frac{\theta_{max} - 2 + \theta_{max}(\theta_{max} - 1)^{D_T} - \theta_{max}}{\theta_{max} - 2} \\
&= \frac{\theta_{max}(\theta_{max} - 1)^{D_T} - 2}{\theta_{max} - 2} \\
&< \frac{\theta_{max}(\theta_{max} - 1)^{D_T}}{\theta_{max}} \qquad \text{(assuming } \theta_{max} > 2\text{)} \\
&= (\theta_{max} - 1)^{D_T}.
\end{aligned}$$

As a result, $\theta_{max} \geq 1 + \sqrt[D_T]{n}$.  ∎

We can derive the storage lower bound easily and is stated as follows.

**Theorem 6** *For a maximum acceptable key path length $D$, the minimum key storage $\theta_{min}(n, D)$ needed at each node for a sensor network with $n$ nodes is given by:*

$$\theta_{min}(n, D) \geq 1 + \sqrt[D]{n}.$$

In order to achieve resiliency against dead nodes and links which happen right after the sensor network is deployed, additional key storage provision is usually needed to allow alternative choices for the acquaintance chain between any two sensor nodes. Doing a similar analysis, we can arrive at the following corollary for a t-graph wherein any pair of non-adjacent vertices are connected by at least $f$ disjoint paths.

**Corollary 7** *Given a t-graph $G_T$ having $n$ vertices with diameter $D_T$ and there are at least $f$ disjoint paths between any two non-adjacent vertices, for the maximum node degree $\theta_{max} > 2$, the following holds:*

$$\theta_{max} \geq 1 + \sqrt[D_T]{fn}$$

The above theorem in essence tells us that if we want to ensure that any two sensor nodes could authenticate each other through another node, we need at least a square root storage (with respect to the total number of nodes) at each sensor node. If we could tolerate a longer acquaintance chain, say 3 hops, we only need a cube-root storage per node. Node that this discussion does not take into account of the deleted edges or vertices in the actual sensor network, but combined with the results of the last section and the following theorem on average distance, we can determine the minimum $\theta_{max}$ needed to meet a given $\overline{W}$ constraint.

**Theorem 8** *Given a t-graph $G_T$ having $n$ vertices with diameter $D_T$, maximum degree $\theta_{max} > 2$ and minimum degree $\theta_{min} > 2$, the following holds for the mean distance $\overline{d}_T$:*

$$d_L < \overline{d}_T < d_U$$

where $d_L = D_T - \frac{\theta_{max}}{(n-1)(\theta_{max}-2)^2}(\theta_{max} - 1)^{D_T}$
and $d_U = D_T - \frac{1}{(n-1)(\theta_{min}-2)}[(\theta_{min} - 1)^{D_T} - (D_T + 1)]$.

**Proof** Suppose we first consider a regular graph with node degree $\theta$ and diameter $D$. Picking any node $s_i$, the number of nodes at 1 hop away is $\theta$, at $i$ hops away is $\theta(\theta - 1)^{i-1}$. We could take the number of nodes at $D$ hops away to be $(n - 1)$ minus the sum of all these. The mean distance $\overline{d}(\theta)$ for this regular graph could be computed as follows:

$$\begin{aligned}
(n-1)\overline{d}(\theta) &= 1 \cdot \theta + 2 \cdot \theta(\theta - 1) + \ldots + (D - 1) \cdot \theta(\theta - 1)^{D-2} \\
&\quad + D \cdot [(n - 1) - (\theta + \theta(\theta - 1) + \ldots + \theta(\theta - 1)^{D-2})] \\
&= \tfrac{\theta}{\theta-1}\sum_{i=1}^{D-1} i(\theta - 1)^{i-1} + (n - 1)D - \tfrac{\theta}{\theta-1}\sum_{i=1}^{D-1}(\theta - 1)^i \\
&= D(n - 1) + \tfrac{\theta}{\theta-1}\tfrac{(D-1)(\theta-1)^{D+1}-D(\theta-1)^D+(\theta-1)}{(\theta-2)^2} - D\tfrac{\theta}{\theta-1}\tfrac{(\theta-1)^D-1}{(\theta-2)} \\
&\vdots \\
&= D(n - 1) - \tfrac{\theta}{(\theta-1)(\theta-2)^2}[(\theta - 1)^{D+1} - (D + 1)(\theta - 1) + D]
\end{aligned}$$

So, we have

$$\overline{d}(\theta) = D - \frac{\theta \cdot [(\theta - 1)^{D+1} - (D + 1)(\theta - 1) + D]}{(n - 1)(\theta - 1)(\theta - 2)^2}.$$

For a general graph, the mean distance $\overline{d}_T$ should be smaller than $\overline{d}(\theta_{min})$. That is,

$$\begin{aligned}
\overline{d}_T &\leq D_T - \tfrac{\theta_{min}[(\theta_{min}-1)^{D_T+1}-(D_T+1)(\theta_{min}-1)+D_T]}{(n-1)(\theta_{min}-1)(\theta_{min}-2)^2} \\
&< D_T - \tfrac{\theta_{min}[(\theta_{min}-1)^{D_T+1}-(D_T+1)(\theta_{min}-1)]}{(n-1)(\theta_{min}-1)(\theta_{min}-2)^2} \\
&= D_T - \tfrac{\theta_{min}[(\theta_{min}-1)^{D_T}-(D_T+1)]}{(n-1)(\theta_{min}-2)^2} \\
&< D_T - \tfrac{[(\theta_{min}-1)^{D_T}-(D_T+1)]}{(n-1)(\theta_{min}-2)}.
\end{aligned}$$

Similarly, $\overline{d}_T > \overline{d}(\theta_{max})$, and we have

$$\begin{aligned}
\overline{d}_T &\geq D_T - \frac{\theta_{max}[(\theta_{max}-1)^{D_T+1} - (D_T+1)(\theta_{max}-1) + D_T]}{(n-1)(\theta_{max}-1)(\theta_{max}-2)^2} \\
&> D_T - \frac{\theta_{max}}{(n-1)(\theta_{max}-1)(\theta_{max}-2)^2}(\theta_{max}-1)^{D_T+1} \\
&= D_T - \frac{\theta_{max}}{(n-1)(\theta_{max}-2)^2}(\theta_{max}-1)^{D_T}.
\end{aligned}$$

This concludes the proof. ∎

### 4.1 An Upper Bound on the Probability of Compromise

We have not considered trading off security for reduced storage in the discussion so far. When a node is compromised, its keys would not leak out any other keys held by other users except the indirectly established keys through this compromised node and the nodes in question could always establish another indirect key through an alternative acquaintance path.

In order to reduce key storage, we could repeat the usage of the keys, say, the a key of a user is shared with a multiple of its neighbors in the t-graph up to a maximum $g$. We could view the key assignment by a labeled or colored t-graph of key graph such that each distinct label or color corresponds to a key. In the case of repeated key usage, a color would appear more than once in the colored t-graph. Hence, when a key is compromised, all edges in the graph with the same label or color would be affected and cannot be used. The following theorem bounds the number of links affected when a node is compromised.

**Theorem 9** *Given a t-graph on $n$ vertices with maximum and minimum vertex degree $\theta_{max}$ and $\theta_{min}$ and diameter $D$. Suppose each node in a sensor network uses a single key for at most $g$ times. The fraction $p_{compromise}$ of links affected is bounded by the following:*

$$p_{compromise} < \frac{g}{g-2} \frac{\theta_{max}}{n\theta_{min}}[(g-1)^D - 1]$$

**Proof** When a node is compromised, at most $\theta_{max}$ neighbors are affected and each of which affects at most $(g-1)$ neighbors. Hence, the total number of nodes affected is bounded by:

$$\begin{aligned}
n' &= \theta_{max} + \theta_{max}(g-1) + \theta_{max}(g-1)^2 + \ldots + \theta_{max}(g-1)^{D-1} \\
&= \frac{\theta_{max}[(g-1)^D - 1]}{g-2}.
\end{aligned}$$

There are at most $g$ edges in each of them are affected. As a result, the maximum number of edges affected is given by $\frac{n'g}{2}$. The total number of edges in a t-graph is at least $\frac{n\theta_{min}}{2}$. As edges and vertices are randomly deleted, we can therefore take this as the fraction $\frac{n'g}{n\theta_{min}}$ over a t-graph as that in the actual network. ∎

Note that Theorem 9 can be easily extended for multiple compromised nodes. A straightforward but less tight bound can be obtained by adding a multiplicative constant.

# 5 Optimal Trust Graph Constructions

To design an optimal key assignment for a given sensor network scenario, we usually need to construct a t-graph of $n$ vertices with minimum vertex degree subject to the following constraints:

- a given diameter $D$;
- a given minimum number of paths between two non-adjacent vertices $f$.

## 5.1 Construction 1: Heuristic Approach

To construct such a graph, we could use a naive heuristic approach as follows: We start with a complete graph $K_n$ on all the $n$ vertices. For each edge in $K_n$, delete it and check whether the constraints are still fulfilled. If yes, update the graph, otherwise add back the edge.

## 5.2 Construction 2: de Bruijn Graph

If we just consider $f = 1$, we could use the de Bruijn graph family [9, 26] and their variations which have very simple and nice structure.

The basic construction consists of the vertex set $\{(a_1, a_2, \ldots, a_p, \ldots, a_r) : a_p \in [1, q]\}$, with $(a_1, a_2, \ldots, a_p, \ldots, a_r)$ adjacent to $(a_2, \ldots, a_p, \ldots, a_r, b)$ for any $b \in [1, q]$. Since the de Bruijn graph is well known, there are various methods to construct it for different number of vertices [7, 14].

Such a graph has degree $\theta = 2q$ and diameter $D = r$ on $n = q^r$ vertices. Note that $\theta = 2\sqrt[D]{n}$ which is just about two times of the lower bound given by Theorem 5.

## 5.3 Construction 3: A variation of de Bruijn Graph

If we do not insist on any value of $f$, the following variation for $f > 1$ could be used. Again, the vertex set is $\{(a_1, a_2, \ldots, a_p, \ldots, a_r) : a_p \in [1, q]\}$ but $q$ has to be prime. Randomly pick $u > r$ vectors from $\mathbb{Z}_q^r$, say $\mathbf{b}_1, \mathbf{b}_2, \ldots, \mathbf{b}_u$, such that any $r$ of these $u$ vectors form a basis of $\mathbb{Z}_q^r$. A vertex $\mathbf{a}$ is made adjacent to $\mathbf{a} + x\mathbf{b}_i$, for all $x \in \mathbb{Z}_q^*$ and $1 \leq i \leq u$.

Such a graph has degree $\theta = u(q-1)$ and diameter $D = r$ on $n = q^r$ vertices and any pair of nodes are connected by $\frac{u!}{(u-r)!}$ paths and $u$ disjoint paths. Note that $\theta \approx f \sqrt[D]{n}$. The reason for this is as follows:

Given any pair of vertices $\mathbf{a}_1$ and $\mathbf{a}_2$, the difference $\mathbf{a}_1 - \mathbf{a}_2$ could always be expressed as a linear combination of any $r$ vectors from $\{\mathbf{b}_1, \mathbf{b}_2, \ldots, \mathbf{b}_u\}$. Suppose we pick $r$ of them and call them $\mathbf{c}_1, \mathbf{c}_2, \ldots, \mathbf{c}_r$. Then, $\mathbf{a}_1 = \mathbf{a}_2 + x_1\mathbf{c}_1 + x_2\mathbf{c}_2 + \ldots + x_r\mathbf{c}_r$ for some $x_1, x_2, \ldots, x_r$. Note that $\mathbf{a}_1$ is adjacent to $\mathbf{a}_1 + x_1\mathbf{c}_1$, which in turn is adjacent to $\mathbf{a}_1 + x_1\mathbf{c}_1 + x_2\mathbf{c}_2$ and so on. In other words, $\mathbf{a}_2$ is reachable from $\mathbf{a}_1$ in at most $r$ hops. Hence, the diameter $D = r$. Since there are $r!$ ways in traversing these $r$ vertices and $\binom{u}{r}$ possible combinations of these $r$ vertices, we have the total number of paths is $\binom{u}{r}r! = \frac{u!}{(u-r)!}$.

# 6  Conclusions

We apply graph theory to study key assignment methods for key predistribution in wireless sensor networks. We map the parameters of a key predistribution scheme with that of a t-graph, which represent the trust relationships between sensor nodes. We give a storage lower bound and an upper bound on compromising probability of key predistribution schemes with a given design constraint of maximum acceptable key path length. We also show a number of near optimal construction from graph theory. We believe better constructions can result via the proposed framework.

# References

[1] S. A. Camtepe and B. Yener. Combinatorial design of key distribution for wireless sensor networks. In *Computer Security - ESORICS, Springer-Verlag LNCS vol. 3193*, pages 293–308, 2004.

[2] S. A. Camtepe and B. Yener. Combinatorial design of key distribution mechanisms for wireless sensor networks. *IEEE/ACM Transactions on Networking*, 15(2):346–358, 2007.

[3] D. W. Carman, P. S. Kruus, and B. J. Matt. Constraints and approaches for distributed sensor network security. *NAI Labs Technical Report #00-010*, September 2000.

[4] H. Chan, A. Perrig, and D. Song. Random key predistribution schemes for sensor networks. In *the Proceedings of IEEE Symposium on Security and Privacy 2003*, pages 197–213, May 2003.

[5] F. R. K. Chung and M. R. Garey. Diameter bounds for altered graphs. *Journal of Graph Theory*, 8:511–534, 1984.

[6] D. E. Clarke, J-E. Elien, C. M. Ellison, M. Fredette, A. Morcos, and R. L. Rivest. Certificate chain discovery in SPKI/SDSI. *Journal of Computer Security*, 9(4):285–322, 2001.

[7] C. J. Colbourn and J. H. Dinitz. *Handbook of Combinatorial Designs*. CRC Press, Inc., Boca Raton, FL, USA, 1996.

[8] D. Cooper, S. Santesson, S. Farrell, S. Boeyen, R. Housley, and W. Polk. Internet X.509 Public Key Infrastructure Certificate and Certificate Revocation List (CRL) Profile. RFC 5280 (Proposed Standard), May 2008.

[9] N. G. de Bruijn. A combinatorial problem. *Koninklijke Nederlandse Akademie v. Wetenschappen*, 49:758–764, 1946.

[10] W. Du, J. Deng, Y. S. Han, S. Chen, and P. Varshney. A key management scheme for wireless sensor networks using deployment knowledge. In *the Proceedings of IEEE INFOCOM'04*, pages 586–597, March 2004.

[11] W. Du, J. Deng, Y. S. Han, and P. Varshney. A pairwise key pre-distribution scheme for wireless sensor networks. In *the Proceedings of ACM Conference on Computer and Communication Security (CCS 2003)*, pages 42–51, November 2003.

[12] W. Du, J. Deng, Y. S. Han, P. K. Varshney, J. Katz, and A. Khalili. A pairwise key predistribution scheme for wireless sensor networks. *ACM Transactions on Information and System Security*, 8(2):228–258, 2005.

[13] L. Eschenauer and V. D. Gilgor. A key management scheme for distributed sensor networks. In *the Proceedings of ACM Conference on Computer and Communication Security (CCS 2002)*, pages 41–47, November 2002.

[14] J. L. Gross and J. Yellen. *Handbook of Graph Theory*. CRC Press, Inc., Boca Raton, FL, USA, 2003.

[15] Q. Jiang, D. S. Reeves, and P. Ning. Certificate recommendations to improve the robustness of web of trust. In *the Proceedings of ISC 2004, Springer-Verlag LNCS vol. 3225*, pages 292–303, 2004.

[16] J. Lee and D. R. Stinson. On the construction of practical key predistribution schemes for distributed sensor networks using combinatorial designs. *ACM Transactions on Information and System Security*, 11(2), 2008.

[17] D. Liu and P. Ning. Establishing pairwise keys in distributed sensor networks. In *the Proceedings of ACM Conference on Computer and Communication Security (CCS 2003)*, pages 52–61, November 2003.

[18] D. Liu, P. Ning, and R. Li. Establishing pairwise keys in distributed sensor networks. *ACM Transactions on Information and System Security*, 8(1):41–77, 2005.

[19] T. Matsumoto and H. Imai. On the key predistribution system: A practical solution to the key distribution problem. In *Advances in Cryptology — CRYPTO 1987, Springer-Verlag LNCS vol. 293*, pages 185–193, 1988.

[20] T. A. McKee and F. R. McMorris. *Intersection Graph Theory*. SIAM, 1999.

[21] A. J. Menezes, P. C. van Oorschot, and S. A. Vanstone. *CRC Handbook of Applied Cryptography*. CRC Press, Inc., Boca Raton, FL, USA, 1996.

[22] C. J. Mitchell and F. C. Piper. Key storage in secure networks. *Discrete Applied Mathematics*, 21:215–228, 1988.

[23] R. Needham and M. Schroeder. Using encryption for authentication in large networks of computers. *Communications of ACM*, 21(12):993–999, 1978.

[24] B. C. Neuman and T. Tso. Kerberos: An authentication service for computer networks. *IEEE Transactions on Communication*, 32:33–38, 1994.

[25] J. Spencer. *The Strange Logic of Random Graphs*. Springer-Verlag, 2000.

[26] F. L. Zhang and G. N. Lin. On the de bruijn-good graphs. *Acta Math. Sinica*, 30(2):195–205, 1987.